\documentclass[twocolumn,twoside,slac_two]{revtex4}
\usepackage{graphicx}
\usepackage{fancyhdr}
\newcommand{\sigmav}{\ensuremath{\langle \sigma v \rangle}}
\newcommand{\clumpi}{\mbox{1FGL~J0030.7+0724}}

\begin{document}

\title{Dark matter subhalos as \textit{Fermi} gamma-ray sources and first candidates in the 1FGL catalog}

\author{H.-S.~Zechlin, M.~V.~Fernandes, D.~Horns}
\affiliation{University of Hamburg, Institut f\"ur Experimentalphysik, Luruper Chaussee 149, D-22761 Hamburg, Germany}
\author{D.~Els\"asser}
\affiliation{University of W\"urzburg, Institut f\"ur Theoretische Physik und Astrophysik, Am Hubland, D-97074 W\"urzburg, Germany}

\begin{abstract}
Predicted by hierarchical structure formation, \mbox{Milky~Way-type}
galaxies are anticipated to host numerous dark matter subhalos with
masses between $10^{10}$ and a cut-off of $10^{-6}\,\mathrm{M}_\odot$,
or even less. In self-annihilating dark matter scenarios, these
objects could be visible in the $\gamma$-ray band as faint and
non-variable sources without astrophysical counterpart. In accordance
with realistic subhalo models and current observational constraints on
self-annihilating dark matter, we predict that about one massive
Galactic subhalo may have already been detected in the 11-months
catalog of \textit{Fermi}-LAT data (1FGL). Selection cuts applied to
the 1FGL reveal twelve possible candidates, and in-depth studies of
the most promising object, \clumpi, are presented. In a dedicated
X-ray follow-up observation with the Swift~XRT, seven point-like X-ray
sources have been discovered. Within the positional uncertainty
derived from the 24-months \textit{Fermi}-LAT data, we consider the
unidentified radio source NVSS~J003119+072456, coinciding with one of
the discovered Swift sources, as the most promising counterpart
candidate for \clumpi. The broad-band spectral energy distribution is
consistent with a high-energy-peaked blazar. However, flux and extent
of the \textit{Fermi} source may also be compatible with a dark matter
subhalo. A discrimination between the two scenarios requires further
multi-wavelength observations. Strategies for identifying $\gamma$-ray
sources associated with self-annihilating DM subhalos are discussed.
\end{abstract}

\maketitle

\thispagestyle{fancy}

\section{Introduction}
Unraveling the nature of dark matter (DM) is part of the major tasks
in modern astro- and particle physics. Various independent
astrophysical observations indicate a non-baryonic form of cold dark
matter (CDM) to prevail over the baryonic content of the Universe,
e.g., \cite{2005PhR...405..279B,2010Natur.468..389B}. Provided by
theories extending the standard model of particle physics (SM), such
as supersymmetry and universal extradimensions, a class of promising
CDM candidates are stable weakly interacting massive particles
(WIMPs). By self-annihilation in SM final states (heavy quarks, gauge
bosons, or leptons), WIMPs can produce detectable signatures such as
photons, antimatter, and leptons, arising from hadronisation and
subsequent decay of the final annihilation states. Apart from
(loop-suppressed) line contributions, these processes result in
continuous photon spectra, following a hard power law (index $\Gamma
\lesssim 1.5$) cutting off exponentially to the WIMP mass.

Within the framework of hierarchical structure formation, DM halos of
\mbox{Milky~Way-type} galaxies should contain numerous DM subhalos
orbiting around the center, with masses between a cut-off scale
$10^{-11}-10^{-3}$ and $10^{10}\,\mathrm{M}_\odot$
\cite{2009NJPh...11j5027B}, where $\mathrm{M}_\odot$ denotes the solar
mass unit. Within the resolved mass scales, numerical high resolution
$N$-body simulations of structure formation, such as the Aquarius
Project \cite{2008Natur.456...73S} or the Via Lactea II simulation
\cite{2008Natur.454..735D}, predict the subhalos to follow a power-law
distribution in mass, $\mathrm{d}N/\mathrm{d}M \propto M^{-\alpha}$,
where $\alpha \in [1.9;2.0]$; their density profiles resemble the host
halo's, resulting in high central densities. Spatially, subhalos
follow an ``anti-biased'' distribution, i.e., the majority orbits far
away from the host's center.

With lacking baryonic but dominating WIMP content, subhalos could
appear as non-variable, faint point-like or moderately extended
high-energy (HE) $\gamma$-ray sources without astrophysical
counterpart in any other wavelength band. A small fraction could then
be detectable with current high- or very-high-energy (VHE)
$\gamma$-ray telescopes
\cite{2008MNRAS.384.1627P,2008ApJ...686..262K,2009PhRvD..80b3520A,2010PhRvD..82f3501B}
such as \textit{Fermi}-LAT ($20\,\mathrm{MeV} - 300\,\mathrm{GeV}$)
\cite{2009ApJ...697.1071A} and imaging air Cherenkov telescopes
(IACTs; $E\gtrsim 100$\,GeV) \cite{2009ARA&A..47..523H}.

This article summarizes the main results of a study on the
detectability of DM subhalos with \textit{Fermi}-LAT, their
properties, and the investigation of the most promising object from
first searches for subhalo candidates in the first \textit{Fermi}-LAT
point-source catalog (1FGL), \clumpi. For details and further
explanation the reader is referred to Zechlin et al., 2011
\cite{2011AA...submittedZ}.

\section{\label{sect:gamma_rays}Gamma rays from DM subhalos}
The total rate of photons with energy $E$ in the intervall
$[E_1;E_2]$, originating from self-annihilating DM in a DM subhalo, is
\begin{equation} \label{eq:luminosity}
 \mathcal{L} = \frac{\sigmav_\mathrm{eff} N_\gamma}{2 m_\chi^2} \int \mathrm{d}V \rho^2(r), \,\,\, N_\gamma = \hskip -1mm \int\limits_{E_1/m_\chi}^{E_2/m_\chi} \mathrm{d}x \frac{\mathrm{d}N_\gamma}{\mathrm{d}x},
\end{equation}
where $\sigmav_\mathrm{eff}$ is the thermally-averaged annihilation
cross section times the relative velocity, $m_\chi$ the WIMP mass, and
$\mathrm{d}N_\gamma/\mathrm{d}x$, $x \equiv E/m_\chi$, denotes the
differential spectrum of photons per annihilation
\cite{2004PhRvD..70j3529F}. The produced photon flux is
$\phi=\mathcal{L}/(4\pi D^2)$, assuming $r_\mathrm{s} \ll D$ (see
below). Consistent with numerical simulations, the density profile of
DM subhalos is taken to follow a Navarro-Frenk-White (NFW) profile
\cite{1997ApJ...490..493N},
scaled to the characteristic inner radius $r_\mathrm{s}$ and
characteristic inner density $\rho_s$. For subhalos, where (e.g.,
tidal) disturbances caused by their formation history are negligible,
the profile parameters are related by the virial halo mass
$M_\mathrm{vir}$. The quantity $M_\mathrm{vir}$ is given by the mass
inside the sphere of radius $R_{\mathrm{vir}}$, which encloses a mean
density of $\Delta_{\mathrm{c}}$ times the critical density of the
Universe $\rho_\mathrm{crit}$ at the considered redshift $z$
\cite{1997ApJ...490..493N}, $M_\mathrm{vir} := 4\pi/3
\,\Delta_\mathrm{c} \rho_\mathrm{crit} R^3_\mathrm{vir}$. The virial
overdensity at $z = 0$ is $\Delta_\mathrm{c} \approx 100$
\cite{1998ApJ...495...80B}. Integrating the NFW profile, the subhalo
mass $M$ is given by $M = 4\pi \rho_\mathrm{s} r^3_\mathrm{s} f(c)$,
where $f(c) \equiv \ln(1+c)-c/(1+c)$ and $c$ denotes the concentration
parameter of the subhalo. For non-disturbed halos, the concentration
is defined by $c_\mathrm{vir} \equiv R_\mathrm{vir}/r_\mathrm{s}$,
depending on the subhalo mass and redshift. In the following, we use
the low-mass extrapolation of a toy model for $c_\mathrm{vir}$
\cite{2008A&A...479..427L}, based on \cite{2001MNRAS.321..559B}.
Corrected by a factor depending on the galactocentric distance,
effects of subhalo formation processes will be included. In this
model, henceforth \textit{subhalo model} (SHM), the concentration
increases with decreasing galactocentric distance, as indicated by
numerical simulations \cite{2008ApJ...686..262K}. Intrinsic to the
stochastic process of halo formation, the concentration of individual
subhalos scatters around the median, see
\cite{2001MNRAS.321..559B,2002ApJ...568...52W}. A general study of the
concentration-model dependence and further details are given in
\cite{2011AA...submittedZ}.

Given the relations above, Eq. \ref{eq:luminosity} simplifies to
\begin{equation} \label{eq:luminosity_vir}
 \mathcal{L} = \frac{\sigmav_\mathrm{eff} N_\gamma \Delta_\mathrm{c}
   \rho_\mathrm{crit}}{18 m_\chi^2} \frac{ M
   c_\mathrm{vir}^3}{f(c_\mathrm{vir})^2}.
\end{equation}
Conveniently, $\sigmav_\mathrm{eff}$ is normalized to $\sigmav_0 = 3
\times 10^{-26}\,\mathrm{cm}^3 \mathrm{s}^{-1}$, which coincides with
the correct relic density. A higher annihilation rate, a so-called
boost factor $\sigmav_\mathrm{eff}/\sigmav_0$, could in general be
related to sub-substructures or the particular particle physics
framework.

\section{\label{sect:Fermi_subhalos}\textit{Fermi}-LAT sources as DM subhalos}
Based on the observational quantities flux and angular extent,
candidate sources for DM subhalos will be selected. With
Eq. \ref{eq:luminosity_vir}, the effective self-annihilation cross
section $\sigmav_\mathrm{eff}$ required for a given flux $\phi$ and
intrinsic source extent $\theta_\mathrm{s}$ is determined with
$\mathcal{L}=4\pi D^2 \phi$, where $\theta_\mathrm{s}$ constrains the
distance $D$ to the subhalo. The intrinsic extent of a subhalo is
traced by the characteristic profile radius $r_\mathrm{s}$, since
$87.5\%$ of the total $\gamma$-ray luminosity is produced within
$r_\mathrm{s}$. Hence, $D \approx r_\mathrm{s}/\theta_\mathrm{s}$,
where $\theta_\mathrm{s}$ denotes the angle corresponding to
$r_\mathrm{s}$. About 68\% of the $\gamma$-ray luminosity is emitted
within $\theta_{68} \simeq 0.46\,\theta_\mathrm{s}$. The relations
below are given with respect to $\theta_\mathrm{s}$, but can easily be
adapted for $\theta_{68}$, which represents the extent for comparision
with observational data. With the WIMP model, $\sigmav_\mathrm{eff}$
is fully determined by the subhalo mass and the observed flux and
extent:
\begin{equation} \label{eq:boost}
 \sigmav_\mathrm{eff} = 96\,\pi^{\frac{1}{3}}
 \frac{m^2_\chi}{N_\gamma} \left( \frac{3}{4 \Delta_\mathrm{c}
   \rho_\mathrm{crit}} \right)^{5/3} \frac{\phi}{\theta^2_\mathrm{s}}
 \, \frac{M^{-1/3} f(c_\mathrm{vir})^2}{c^5_\mathrm{vir}}.
\end{equation}
For heavy WIMPs, a detection of DM subhalos is favored in the
high-energy band of \textit{Fermi}-LAT ($10-100\,\mathrm{GeV}$), due
to improving sensitivity \citep{2009ApJ...697.1071A} and the energy
spectrum of DM annihilation. As will be demonstrated later, a fiducial
subhalo candidate source is characterized by faintness and a moderate
angular extent. Therefore, the flux of the fiducial source is chosen
to be at the level of detection sensitivity (one year), which has been
accurately studied in \cite{2011AA...submittedZ}:
$\phi^\mathrm{fid}(10 - 100\,\mathrm{GeV}) = 1.6 \times
10^{-10}\,\mathrm{cm}^{-2}\,\mathrm{s}^{-1}$ for
$\theta^\mathrm{fid}_\mathrm{s} = 1^\circ$, corresponding to
$\theta^\mathrm{fid}_{68} \approx 0.5^\circ$; the Galactic position
has been chosen to match \clumpi.

\begin{figure}[t]
\resizebox{\hsize}{!}{\includegraphics{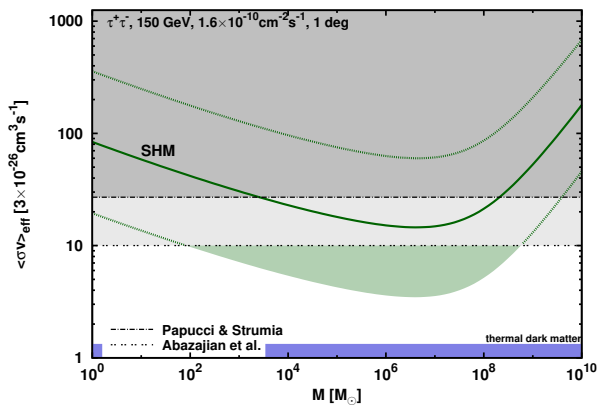}}
\caption{Boost factor $\sigmav_\mathrm{eff}$ required for the fiducial
  \textit{Fermi}-LAT source to originate from a DM subhalo of mass
  $M$. The solid green line indicates the prediction of the average
  SHM model, while its scatter is depicted by the dotted green lines
  (and the green-shaded area). Current observational constraints on
  $\sigmav_\mathrm{eff}$ are shown by the light grey- and grey-shaded
  area.}
\label{fig:fiducial}
\end{figure}

For WIMPs of $m_\chi = 150\,\mathrm{GeV}$ fully annihilating to
$\tau^+\tau^-$, the effective boost factors
$\sigmav_\mathrm{eff}/\sigmav_0$ required to generate the emission
$\phi^\mathrm{fid}$ of the fiducial source by DM annihilation are
depicted in Fig. \ref{fig:fiducial}. Within the halo-to-halo scatter
of the concentration, the required boost spans about one order of
magnitude. Minimal boost is required for massive subhalos between
$10^{6}$ and $10^{7}\,\mathrm{M}_\odot$. In Fig. \ref{fig:fiducial},
the necessary effective annihilation cross section is compared with
current observational limits on $\sigmav_\mathrm{eff}/\sigmav_0$
\cite{2010JCAP...11..041A,2010JCAP...03..014P}. The Figure shows that,
within the scatter, the boost factor needed to explain $\gamma$-ray
sources such as the fiducial with massive DM subhalos ($\sim 10^5 -
10^8\,\mathrm{M}_\odot$) is consistent with observational
constraints. These subhalos are in corresponding distances from 0.5 to
10\,kpc. Further reduction of the boost could be accomplished by
additional sub-substructure (factor of 2-3,
\cite{2009JCAP...06..014M}). For the considered WIMP model, this leads
to a required boost of order unity within the scatter.

For this WIMP model and current observational constraints on
$\sigmav_\mathrm{eff}$, on average 0.2 subhalos between $10^5$ and
$10^8\,\mathrm{M}_\odot$ are expected for detection with
\textit{Fermi}-LAT in one year, see \cite{2011AA...submittedZ}. Given
the Poisson distribution of the average, up to one massive subhalo is
expected in the one-year data set.

\section{Searches for DM subhalos in the 1FGL}
As shown above, DM subhalos would appear as faint, non-variable, and
moderately extended sources without astrophysical counterparts. First
candidates have been searched for in the 11-months point-source
catalog of \textit{Fermi}-LAT (1FGL, 100\,MeV - 100\,GeV,
\cite{2010ApJS..188..405A}), listing 1451 $\gamma$-ray sources
together with flux, position, significance of variability, and
spectral curvature. Among the sources, 630 objects are not confidently
associated with known sources at other wavelengths.

To select possible subhalo candidates, the sample of unassociated
sources has been scanned for non-variable sources detected between 10
and 100\,GeV. A high-energy detection assures subhalo candidates
driven by heavy WIMPs and furthermore avoids confusion with
high-energy pulsars \cite{2007ApJ...659L.125B}. The position in the
Galaxy has been constrained to galactic latitudes $|b|\geq 20^\circ$,
eliminating confusion with Galactic sources.

\textit{Twelve} unassociated sources pass these cuts. Apart from one,
the sample consists of sources at the faint end of the entire 1FGL
sample. A further discussion of these objects is presented in
\cite{2011AA...submittedZ}. According to the 1FGL catalog, the
counterpart search has been extended to a wider choice of astronomical
catalogs. Given the twelve candidates, the most promising has been
selected, governed by lacking association, faintness, and spectral
shape, namely \clumpi.

Both the high-energy flux of \clumpi,
$\phi_\mathrm{p}(10-100\,\mathrm{GeV}) = (1.5 \pm 0.7) \times
10^{-10}\,\mathrm{cm}^{-2}\,\mathrm{s}^{-1}$, as well as the spectral
index, $\Gamma = 1.7 \pm 0.4$, are well compatible with a
self-annihilating DM origin. Note that the source has only been
significantly detected between 10 and 100\,GeV. Within the positional
uncertainty of the $\gamma$-ray signal (at 95\% confidence),
\clumpi\space has no convincing multi-wavelength
counterpart. Furthermore, X-ray observations with ROSAT
\cite{1999A&A...349..389V} constrain the energy-flux of a possible
X-ray counterpart to \mbox{$\lesssim
  10^{-12}\,\mathrm{erg}\,\mathrm{cm}^{-2}\,\mathrm{s}^{-1}$}
\cite{2010BT...thesisB}.

\subsection{\label{sect:Fermi_data}\textit{Fermi}-LAT data}
For further investigation of possible counterparts, temporal
variability, and the angular extent of \clumpi, updated results based
on the 24-months public archival data set of \textit{Fermi}-LAT
between 10 and 100\,GeV have been derived. The point-source analysis
has been performed with the \textit{Fermi Science Tools} v9r18p6
\cite{2011FSSC...webF}, using recommended options and the
instrument-response functions \textit{P6\_V3\_DIFFUSE}
\cite{2009arXiv0907.0626R}. Details are given in
\cite{2011AA...submittedZ}.

Within a radius of 0.5$^\circ$ around the nominal source position, six
photons (one class 3 event, five class 4 events) between 10 and
100\,GeV have been detected after 24 months. With respect to the
signal, the influence of the Galactic foreground as well as the
extragalactic background at the source position is negligible.  The
integrated flux between 10 and 100\,GeV, reconstructed with the
point-source analysis, is $\phi_\mathrm{p}(10-100\,\mathrm{GeV}) =
(0.9\pm 0.4) \times 10^{-10}\,\mathrm{cm}^{-2}\,\mathrm{s}^{-1}$. The
significance of the signal in this energy-bin is about 6.6 Gaussian
standard deviations.

The variability of \clumpi\space has been tested for, analyzing the
temporal photon distribution for compatibility with a constant flux
with an unbinned Kolmogorov-Smirnov test. The test confirms the
null-hypothesis of a steady flux with a probability of about
0.5. Given the positional distribution of the signal photons, the
(intrinsic) spatial extent has been tested with a likelihood-ratio
test, comparing the null-hypothesis of a point-source with the
assumption of the intensity distributed following the line-of-sight
integral of the squared NFW profile. The data favors a moderate extent
$\theta_\mathrm{s} = 0.14^{+0.20}_{-0.12}\,\mathrm{deg}$, which is,
however, not significantly different from a point-source
hypothesis. The upper limit is $\theta_\mathrm{s} \leq
0.72\,\mathrm{deg}$ at 95\% confidence level.

\subsection{\textit{Swift}-XRT data}
New observations of the field with the X-ray telescope (XRT,
0.2-10\,keV) onboard the \textit{Swift} satellite
\cite{2004ApJ...611.1005G} with a total effective exposure of 10.1\,ks
led to the discovery of seven new X-ray sources. The (unabsorbed)
energy fluxes of the discovered sources range from $\sim 2 \times
10^{-14}\,\mathrm{erg}\,\mathrm{cm}^{-2}\,\mathrm{s}^{-1}$ to $2
\times 10^{-13}\,\mathrm{erg}\,\mathrm{cm}^{-2}\,\mathrm{s}^{-1}$
between 0.2 and 2\,keV. See \cite{2011AA...submittedZ} for further
details.

\section{Discussion}
\subsection{\clumpi\space as AGN}
Within the updated positional information on \clumpi\space (see
Sect. \ref{sect:Fermi_data}), the radio source NVSS~J003119+072456
($f_{1.4\,\mathrm{GHz}} = (11.6\pm0.6$)\,mJy) provides a viable
counterpart. The radio source positionally coincides with the newly
discovered hard X-ray source SWIFT~J003119.8+072454 ($\Gamma = 1.6 \pm
0.3$), energy flux \mbox{$\sim 2\times
  10^{-13}\,\mathrm{erg}\,\mathrm{cm}^{-2}\,\mathrm{s}^{-1}$} between
0.2 and 2 keV, and the optical counterpart SDSS~J003119.71+072453.5
($m(r) = 17.4^\mathrm{m}$). Normalized to the radio flux, the average
spectral energy distribution of a high-energy-peaked blazar (HBL, see
\cite{2001A&A...375..739D}) fits the data, assuming standard temporal
variability of the source. Reinforcing, the HBL scenario is favored by
the spectral indices of the X- and $\gamma$-ray signal.

The optical counterpart appears point-like with $m(R) =
18.6^\mathrm{m}$ (the SDSS source coincides with USNO
0974-0005617). Under the assumption of an elliptical host galaxy with
$M_R = -23.1^\mathrm{m}$, typical for blazars
\cite{2007A&A...475..199N}, the corresponding distance modulus implies
a redshift of $z = 0.39 \pm 0.03$ (K-correction).

\subsection{\clumpi\space as DM subhalo}
Given the absence of clear indication for variability, a DM subhalo
origin of \clumpi\space remains plausible. The
flux$^\textnormal{\scriptsize \cite{footnote1}}$ and the upper limits
on the angular extent provided in Sect. \ref{sect:Fermi_data} are
consistent with a subhalo of mass between $10^6$ and
$10^8\,\mathrm{M}_\odot$. Assuming a mass of $10^6\,\mathrm{M}_\odot$,
the distance would be $2.4_{-0.7}^{+1.0}\,\mathrm{kpc}$ given the
scatter of the concentration model, see Sects. \ref{sect:gamma_rays}
and \ref{sect:Fermi_subhalos}. In case of a WIMP of 150\,GeV
annihilating to $\tau^+\tau^-$, the required minimal boost is 3 for a
high-concentrated subhalo with a corresponding distance of
$1.7\,\mathrm{kpc}$, and 13 for an average-concentrated subhalo with a
corresponding distance of $2.4\,\mathrm{kpc}$. An even lower boost may
be required, if the subhalo contains sub-substructures or exhibits a
cuspier profile.

\section{Summary and Outlook}
We have presented a state-of-the-art investigation on the
detectability of DM subhalos with the \textit{Fermi}-LAT. Concluding,
one massive DM subhalo will probably appear in the first
catalog. First searches have indeed revealed twelve possible
candidates, where the most promising, \clumpi, has been studied in
detail. A detection of temporal variability and improved astrometry
would clarify the physical origin of this object.

The accompanying paper Zechlin et al., 2011 \cite{2011AA...submittedZ}
will contain additional details. The recent release of the second
\textit{Fermi}-LAT catalog, 2FGL, allows a deeper search for subhalo
candidates, which will be presented in a future work. Note that
related work is presented in \cite{2011theseprocD} and
\cite{2011arXiv1110.4744N}.

\bigskip 
\begin{acknowledgments}
\footnotesize{We thank our colleagues K. Borm, T. Bringmann,
  W. Buchm\"uller, F. J\"ager, A. Lobanov, and M. Raue for helpful
  discussions. We thank the \textit{Swift}-PI N. Gehrels and his team
  for the observations, following-up on our ToO request. The help of
  the Fermi HelpDesk is acknowledged. This work was supported by the
  collaborative research center (SFB) 676 ``Particles, Strings, and
  the Early Universe'' at the University of Hamburg and the German
  federal ministry for education and research (Bundesministerium f\"ur
  Bildung und Forschung, BMBF).}
\end{acknowledgments}
\vspace{-0.5cm}

\end{document}